\documentclass[12pt]{article} 
\usepackage{simpleConference}
\usepackage{times}
\usepackage{array,multirow,graphicx}
\usepackage{float}
\usepackage{dcolumn,tabularx,booktabs,tabulary}
\usepackage{multirow}
\usepackage{amssymb}
\usepackage{url,hyperref}
\usepackage{authblk}
\usepackage{amsmath}
\usepackage[justification=centering]{caption}
\hypersetup{
    colorlinks=true,
    linkcolor=blue,
    filecolor=magenta,      
    urlcolor=cyan,
}

\usepackage[left=2.54cm, right=3.18cm, top=2.54cm, bottom=2.54cm]{geometry}

\begin{document}

\title{Using Deep Learning Neural Networks and Candlestick Chart Representation to Predict Stock Market}

\author[1]{Rosdyana Mangir Irawan Kusuma}
\author[2]{Trang-Thi Ho}
\author[3]{Wei-Chun Kao}
\author[1]{Yu-Yen Ou}
\author[2]{Kai-Lung Hua}
\affil[1]{Department of Computer Science and Engineering, Yuan Ze University, Taiwan Roc}
\affil[2]{Department of Computer Science and Engineering, National Taiwan University of Science and Technology, Taiwan Roc}
\affil[3]{Omniscient Cloud Technology}
\renewcommand\Authands{ and }

\maketitle

\thispagestyle{empty}

\begin{abstract}Stock market prediction is still a challenging problem because there are many factors effect to the stock market price such as company news and performance, industry performance, investor sentiment, social media sentiment and economic factors. This work explores the predictability in the stock market using Deep Convolutional Network and candlestick charts. The outcome is utilized to design a decision support framework that can be used by traders to provide suggested indications of future stock price direction. We perform this work using various types of neural networks like convolutional neural network, residual network and visual geometry group network. From stock market historical data, we converted it to candlestick charts. Finally, these candlestick charts will be feed as input for training a Convolutional Neural Network model. This Convolutional Neural Network model will help us to analyze the patterns inside the candlestick chart and predict the future movements of stock market. The effectiveness of our method is evaluated in stock market prediction with a promising results 92.2 \% and 92.1 \% accuracy for Taiwan and Indonesian stock market dataset respectively. The constructed model have been implemented as a web-based system freely available at \url{http://140.138.155.216/deepcandle/} for predicting stock market using candlestick chart and deep learning neural networks.
\end{abstract}


\section{Introduction}
The stock market is something that cannot be separated from modern human life. The Investment in stock market is a natural thing done by people around the world. They set aside their income to try their luck by investing in stock market to generate more profit. Traders are more likely to buy a stock whose value is expected to increase in the future. On the other hand, traders are likely to refrain from buying a stock whose value is expected to fall in the future. Therefore, an accurate prediction for the trends in the stock market prices in order to maximize capital gain and minimize loss is urgent demand. Besides, stock market prediction is still a challenging problem because there are many factors effect to the stock market price such as company news and performance, industry performance, investor sentiment, social media sentiment and economic factors. According to Fama's efficient market hypothesis argued that it is impossible for investors to get advantage by buying underrated stocks or selling stocks for exaggerated price\cite{malkiel1970efficient}. Therefore, the investor just has only one way to obtain higher profits is by chance or purchasing riskier investments. With the current technological advances, machine learning is a breakthrough in aspects of human life today and deep neural network has shown potential in many research fields. In this research, we apply different types of machine learning algorithms to enhance our performance result for stock market prediction using convolutional neural network, residual network, virtual geometry group network, k-nearest neighborhood and random forest.

Dataset format in machine learning can be different. Many kind of dataset format such as text sequence, image, audio, video, from 1D (one dimension) to 3D (three dimension) can be applicable for machine learning. Taken as an example, the image is used not only as input for image classification, but also as an input to predict a condition. We take the example of Google DeepMind's research in Alpha Go\cite{he2016deep}. Recently, they are successfully get a lot of attention in the research field. By using the image as their input, where the image represents a Go game board, which later this image dataset is used to predict the next step of the opponent in the Go game. On the other occasion, from historical data of stock market converted into audio wavelength using deep convolutional wave net architecture can be applied to forecast the stock market movement\cite{borovykhdilated}.

Our proposed method in this work is using the represented candlestick charts of Taiwan and Indonesian stock markets to predict the price movement. We utilized  three trading period times to analyze the correlation between those period times with the stock market movement. Our proposed candlestick chart will represent the sequence of time series with and without the daily volume stock data. The experiments in this work conduct two kind of image sizes (i.e. 50 and 20 dimension) for candlestick chart to analyze the correlation of hidden pattern in various image size. Thereafter our dataset will be feed as input for several learning algorithms of random forest and k-nearest neighborhood as traditional machine learning, CNN, residual network and VGG network as our modern machine learning. The goal is to analyze the correlation of some parameters such as period time, image size, feature set with the movement of stock market to check whether it will be going up or going down in the next day.
\section{Related Work}
There are many researchers have been started to develop the computational tool for the stock market prediction. In 1990, Schneburg conducted a study using data from a randomly selected German stock market, then using the back-propagation method for their machine learning architecture \cite{schoneburg1990stock}. To our knowledge, stock market data consist of open price data, close price data, high price data, low price data and volume of the daily movement activity. In addition, to use the historical time series data from the stock market, some researchers in this field of stock market predictions began to penetrate the method of sentiment analysis to predict and analyze movements in the stock market. J. Bollen reported the sentiment analysis method by taking data from one of the famous microblogging site Twitter to predict the Dow Jones Industrial Average $($DJIA$)$ stock market movements\cite{bollen2011twitter}. There are more studies on stock market predictions; they use the input data not only by using elements of historical time series data, but by also processing the data into other different forms. $($Borovykh, Bohte et al.$)$ tried to use the deep convolutional wave net architecture method to perform analysis and prediction using data from S \& P500 and CBOE \cite{borovykhdilated}.
\par
We also found some related works using candlestick charts in their research. (do Prado, Ferneda et al. 2013) used the candlestick chart to learn the pattern contained in Brazilian stock market by using sixteen candlestick patterns\cite{do2013effectiveness}. $($Tsai and Quan 2014$)$ utilized the candlestick chart to combine with seven different wavelet-based textures to analyze the candlestick chart\cite{tsai2014stock}. While,  $($Hu, Hu et al. 2017$)$ used the candlestick chart to build a decision-making system in stock market investment. They used the convolutional encoder to learn the patterns contained in the candlestick chart\cite{hu2017deep} while 
$($Patel, Shah et al. 2015$)$ used ten technical parameters from stock trading data for their input data and compare four prediction models, Artificial Neural Network $($ANN$)$, Support Vector Machine $($SVM$)$, random forest and naïve-Bayes\cite{patel2015predicting}. Traditional machine learning like Random Forest has been applied to predict the stock market with a good result. $($Khaidem, Saha et al. 2016$)$ combine the Random Forest with technical indicator such as Relative Strength Index $($RSI$)$ shown a good performance\cite{khaidem2016predicting}. Adding more feature set can be one of the way to enrich your dataset and enhance the result of classification. According to $($Zhang, Zhang et al. 2018$)$ input data is not only from historical stock trading data, a financial news and users’ sentiments from social media can be correlated to predict the movement in stock market\cite{zhang2018improving}.
\par
Different from most of existing studies that only consider stock trading data, news events or sentiments in their models, our proposed method utilized a representation of candlestick chart images to analyze and predict the movement of stock market with a novel to compare modern and traditional neural network. 
\section{Dataset}
\subsection{Data Collection}
\begin{table}[]
\centering
\caption{The period time of our dataset, separated between the training, testing and independent data.}
\label{Tab:datasettime}
\scalebox{0.8}{%
\begin{tabular}{|l|c|c|c|c|c|c|}
\hline
\multicolumn{1}{|c|}{\multirow{2}{*}{Stock Data}} & \multicolumn{2}{c|}{Training Data} & \multicolumn{2}{c|}{Testing Data} & \multicolumn{2}{c|}{Independent Data} \\ \cline{2-7} 
\multicolumn{1}{|c|}{}                            & Start            & End             & Start           & End             & Start             & End               \\ \hline
TW50                                              & 2000/01/01       & 2016/12/31      & 2017/01/01      & 2018/06/14      & 2017/01/01        & 2018/06/14        \\ \cline{2-7} 
ID10                                              & 2000/01/01       & 2016/12/31      & 2017/01/01      & 2018/06/14      & 2017/01/01        & 2018/06/14        \\ \hline
\end{tabular}
}
\end{table}
Getting the right data in the right format is very important in machine learning because it will help our learning system go to right way and achieve a good result. We trained and evaluated our model on two different stock markets, i.e. Taiwan and Indonesia. We collected 50 company stock markets for Taiwan and 10 company stock markets for Indonesia based on their growth in technical analysis as a top stock market in both countries.

In this data collection, we use the application program interface $($API$)$ service from Yahoo! Finance to get historical time series data for each stock market. From the period that we have been set in the following Table \ref{Tab:datasettime}, we certainly get some periods of trading day, starting from Monday until Friday is the period of trading day.

Segregation of data based on predetermined time for data training and data testing is important, while some studies make mistakes by scrambling data; this is certainly fatal because of the data, which we use, is time-series.
\subsection{Data Preprocessing}
From historical time series data, we converted it into candlestick chart using library Matplotlib\cite{hunter2007matplotlib}. To analyze the correlation between different period times with the stock market movement, we divided the data used to create candlestick chart based on three period times such as 5 trading days data, 10 trading days data and 20 trading days data. Besides the period time, we also divided our candlestick chart with and without volume indicator. Adding a volume indicator into candlestick chart is one of our approaches to find out correlation between enrich candlestick chart information and prediction result.
\section{Methodology}
\begin{figure}
\centering
  \includegraphics[width=\columnwidth]{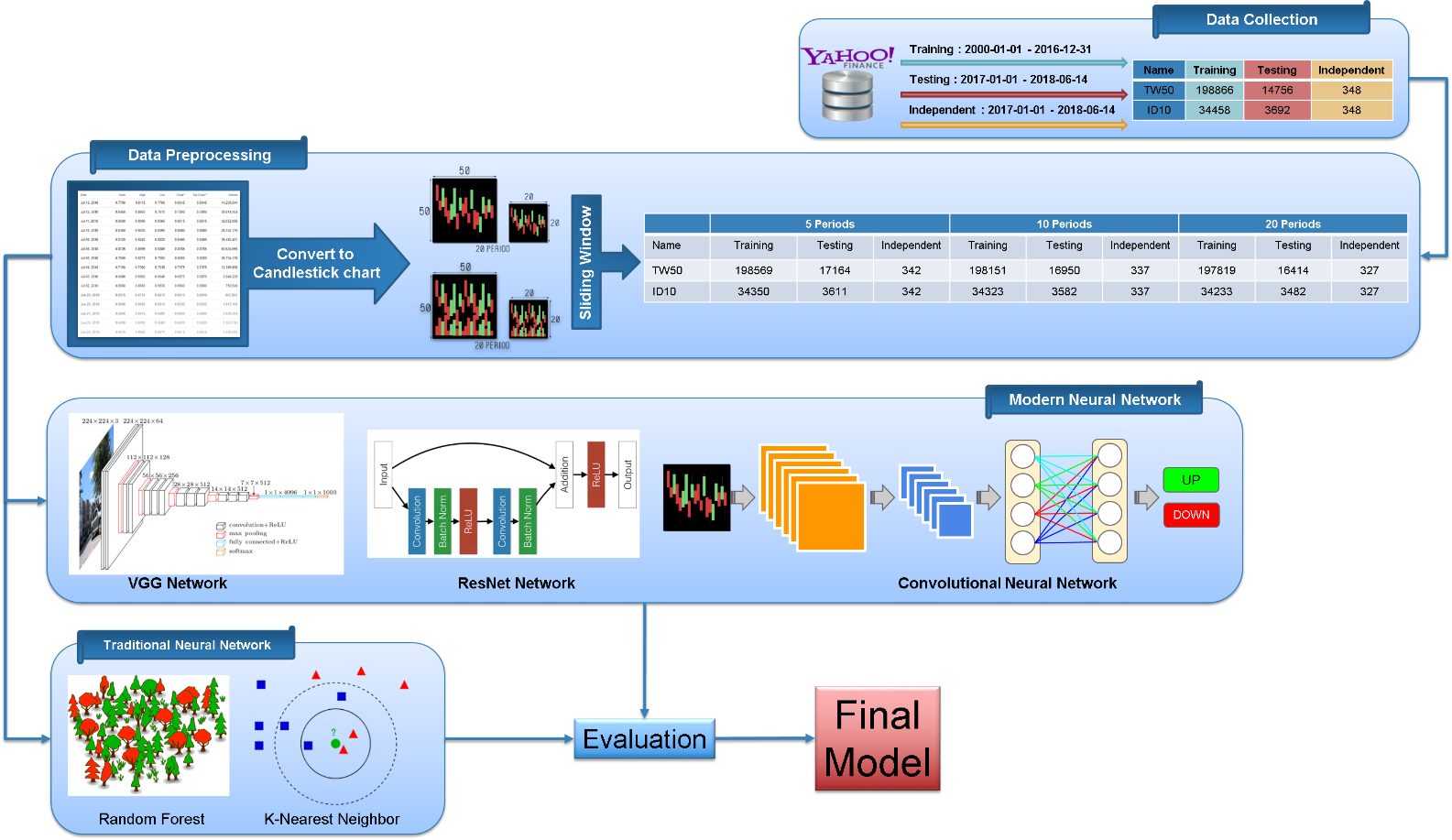}
  \caption{Our methodology design.}
  \label{fig:methodologydesign}
\end{figure}
The architecture of our proposed method is shown in Figure \ref{fig:methodologydesign}. The first, we collected the data from stock market historical data using Yahoo! Finance API. After that, we applied the sliding window technique to generate the period data before using computer graphic technique to generate the candlestick chart images. Finally, our candlestick charts are feed as input into some deep learning neural networks model to find the best model for stock market prediction, and the outputs will be binary class to indicate the stock price will going up or down in the near future.
\subsection{Candlestick Chart}
Candlestick chart is a style of financial chart used to describe the price movements for a given period of time. Candlestick chart is named a Japanese candlestick chart which has been developed by Japanese rice trader- Munehisa Hooma \cite{morris2006candlestick}. Each candlestick typically shows one day of trading data, thus a month chart may show the 20 trading days as 20 candlestick charts. Candlestick chart is like a combination of line-chart and a bar-chart. While each bar represents four important components of information for trading day such as the open, the close, the low and high price.
\begin{figure}
\centering
  \includegraphics[width=0.8\columnwidth]{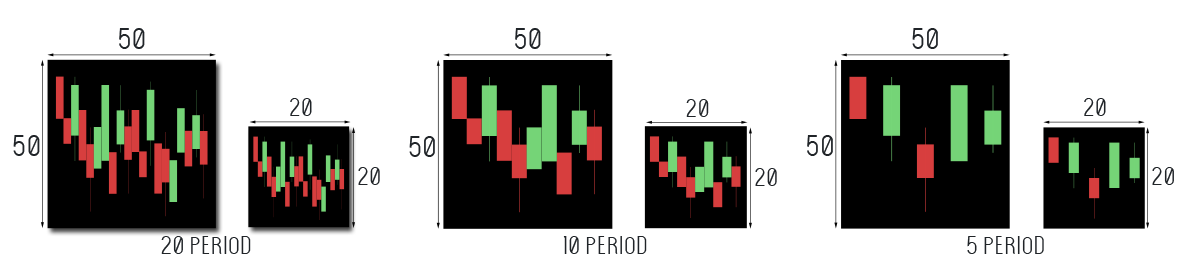}
  \caption{Proposed candlestick chart without volume indicator in different period time and size.}
  \label{fig:candlewithoutvolume}
\end{figure}
\begin{figure}
\centering
  \includegraphics[width=0.8\columnwidth]{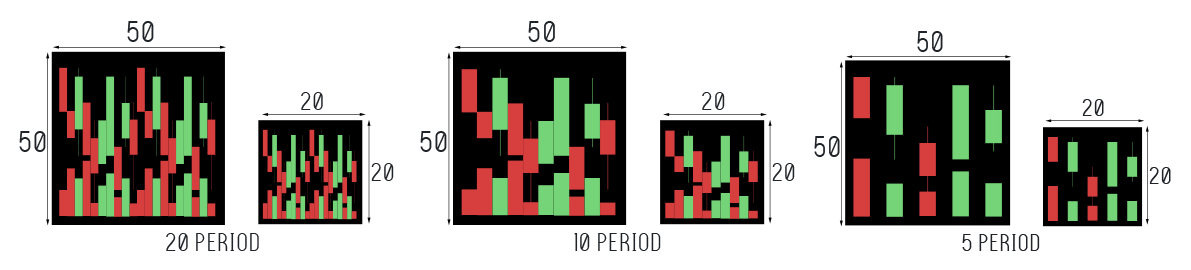}
  \caption{Proposed candlestick chart with volume indicator in different period time and size.}
  \label{fig:candlewithvolume}
\end{figure}
Candlesticks usually are composed of 3 components, such as upper shadow, lower shadow and real body. If the opening price is higher than the closing price, then the real body will filled in red color. Otherwise, the real body will be filler in green color. The upper and a lower shadow represent the high and low price ranges within a specified time period. However, not all candlesticks have a shadow. Candlestick chart is a visual assistance to make a decision in stock exchange. Based on candlestick chart, a trader will be easier to understand the relationship between the high and low as well the open and close. Therefore, the trader can identify the trends of stock market for a specific time frame \cite{lu2012profitable}. The candlestick is called bullish candlestick when the close is greater than the open. Otherwise it is called bearish candlestick.
Figure \ref{fig:candlewithoutvolume} and Figure \ref{fig:candlewithvolume} describe our candlestick chart representation in different period time and size with volume and without volume respectively.

\subsection{Learning Algorithm}
In this work we will use some Deep Learning Networks (DLN) based on Convolutional Neural Network to perform our classification on stock market prediction. Besides the DLN, we also apply some traditional Machine Learning (ML) algorithms to compare with DLN. Those traditional Machine Learning algorithms are Random Forest and K-Nearest Neighbors algorithms.

\subsubsection{Convolutional Neural Network}

 \begin{table}[H]
 \centering
 \caption{Our proposed CNN architecture.}
 \label{Tab:cnnarchitecture}
\scalebox{0.8}{%
\begin{tabular}{|l|l|l|l|l|l|l|l|l|l|l|l|l|l|l|}
\hline
{\rotatebox[origin=c]{90}{Input}} & 
{\rotatebox[origin=c]{90}{ Conv2D-32 ReLU }} & 
{\rotatebox[origin=c]{90}{max-pooling}} & 
{\rotatebox[origin=c]{90}{Conv2D-48 ReLU}} &
{\rotatebox[origin=c]{90}{max-pooling}} & 
{\rotatebox[origin=c]{90}{Dropout}} &
{\rotatebox[origin=c]{90}{Conv2D-64 ReLU}} & 
{\rotatebox[origin=c]{90}{max-pooling}} & 
{\rotatebox[origin=c]{90}{Conv2D-96 ReLU}} & 
{\rotatebox[origin=c]{90}{max-pooling}} & 
{\rotatebox[origin=c]{90}{Dropout}} & 
{\rotatebox[origin=c]{90}{Flatten}} & 
{\rotatebox[origin=c]{90}{Dense-256}} & 
{\rotatebox[origin=c]{90}{Dropout}} & 
{\rotatebox[origin=c]{90}{Dense-2}} \\ \hline
 \end{tabular}
 }
 \end{table}
Convolutional neural network (CNN) is a feed-forward artificial neural networks which includes input layer, output layer and one or more hidden layers. The hidden layers of CNN typically consist of pooling layers, convolution layers and full connected layers. It is similar to ordinary Neural Networks (NN) made up of a set of neurons with learnable weights and bias. The difference is Convolutional layers use a convolution operation to the input then transfer the result to the next layer. This operation allows the forward function more efficient to implement with much fewer parameters.
\par
As shown in Table \ref{Tab:cnnarchitecture} ,our CNN model architecture consist of 4 layers of convolutional 2d, 4 layers of max pooling 2d, and 3 dropouts.
\subsubsection{Residual Network}
It is an artificial neural network developed by He in 2015 \cite{he2016deep}. It uses skip connections or short-cut to jump over some layers. The key of residual network architecture is the residual block, which allows information to be passed directly through. Therefore, the backpropagated error signals is reduced or removed. This allows to train a deeper network with hundreds of layers. And this vastly increased depth led to significant performance achives.
\subsubsection{VGG Network}
The VGG network architecture was introduced by Simonyan and Zisserman\cite{simonyan2014very}. It is named VGG because this architecture is from VGG group, Oxford. This network is characterized by its simplicity, using only 3x3 convolutional layers stacked on top of each other in increasing depth. Reducing volume size is handled by max pooling. Two fully connected layers, each with 4096 nodes are then followed by a softmax classifier. The “16” and “19” stand for the number of weight layers in the network. Unfortunately, there are two major drawbacks with VGGNet. First, it is painfully slow to train and the second the network architecture weights themselves are quite large.
\subsubsection{Random Forest}
Random Forest classifier is a classifier with consist of many decision trees and adopted the technique of random decision forest prioritizes predictive performance by using multiple learning algorithms (ensemble learning). In general, Decision trees are a learning methods used in data search technique. The method used by the idea of combining the "bagging" idea or called "Bootstrap Aggregating" (reduce variance) and the random selection of features in the training sets (classification and regression tree). 
\par
The difference between Random Forest algorithm and the decision tree algorithm is that in Random Forest, the processes of finding the root node and splitting the feature nodes will run randomly. We applied our random forest algorithm from a machine learning python library called skicit-learn\cite{pedregosa2011scikit}.
\subsubsection{K-Nearest Neighbors}
K-Nearest Neighbors (KNN) is a classifier with based on the Lazy learning and Instance-based (IBk) learning algorithms (selection K based value based on model evaluation method or cross validation). Further, Lazy learning is a learning method with the purposed to store training data and enables the training data is used when there is a query request is made (waits until it is given a test) by the system. Similarity measure applied to the KNN with the aim to compare every new case with available cases (training data) that has been previously saved. KNN adopted a supervised learning approach by utilizing the labeled data and this learning model of the algorithm can be used for classification and regression predictive problems.
\par
We also using skicit-learn python library for our KNN classifier. Furthermore, we used a K-D Tree algorithm in our KNN to perform prediction with default parameter from scikit-learn library.
\subsection{Performance Evaluation}
There are some statistics measures of the performance evaluation to evaluate the result of all the classifiers by measuring the sensitivity (true positive rate or recall), specificity (true negative rate), accuracy and Matthew's correlation coefficient (MCC). In general, TP is true positive or correctly identified, FP is false positive or incorrectly identified, TN is true negative or correctly rejected and FN is false negative or incorrectly rejected. Formulated as follows:
\begin{equation}
\small
Sensitivity=\frac{TP}{TP+FN}
\end{equation}
\begin{equation}
\small
Specitivity=\frac{TN}{TN+FP}
\end{equation}
\begin{equation}
\small
Accuracy=\frac{TP+TN}{TP+FP+TN+FN}
\end{equation}
\begin{equation}
\small
MCC=\frac{TP\times TN-FP \times FN}{\sqrt{(TP+FP)(TP+FN)(TN+FP)(TN+FN)}}
\end{equation}
\section{Experimental Results and Discussion}
In this section, we perform classification based on some traditional and modern machine learning algorithms (random forest, kNN, residual network, VGG, CNN) and then evaluate  the  performance  of  our best classification algorithm compared to three state-of-the-art  methods \cite{khaidem2016predicting,patel2015predicting,zhang2018improving}
\subsection{Classification for Taiwan 50 Dataset}
\begin{table}[H]
 \centering
 \caption{Summary result of Taiwan 50 with their best classifier for each trading days and image dimension.}
 \label{Tab:summarytw50}
\scalebox{0.75}{%
 \begin{tabular}{|c|l|c|c|c|c|c|c|c|}
 \hline
 & \multicolumn{1}{c|}{Classifier} & \multicolumn{1}{c|}{Period} & \multicolumn{1}{c|}{Dimension} & \multicolumn{1}{c|}{Sensitivity} & \multicolumn{1}{c|}{Specitivity} & \multicolumn{1}{c|}{Accuracy} & \multicolumn{1}{c|}{MCC}\\
 \hline
 \parbox[t]{2mm}{\multirow{5}{*}{\rotatebox[origin=c]{90}{with volume}}} 
 & CNN & 5 & 50 & 83.2 & 83.8 & 83.5 & 0.67 \\
 & CNN & 10 & 50 & 88.6 & 87.3 & 88.0 & 0.758 \\
 & CNN & 20 & 50 & \textbf{91.6} & \textbf{91.3} & \textbf{91.5} & \textbf{0.827} \\
 & CNN & 5 & 20 & 83.9 & 82.7 & 83.3 & 0.666 \\
 & Random Forest & 10 & 20 & 87.0 & 88.3 & 87.6 & 0.751 \\
 & CNN & 20 & 20 & 90.8 & 90.2 & 90.6 & 0.808 \\
 \hline
  \parbox[t]{2mm}{\multirow{5}{*}{\rotatebox[origin=c]{90}{without volume}}} 
 & CNN & 5 & 50 & 83.6 & 85.1 & 84.4 & 0.687 \\
 & CNN & 10 & 50 & 89.2 & 88.1 & 88.7 & 0.773 \\
 & CNN & 20 & 50 & \textbf{93.3} & 90.7 & \textbf{92.2} & \textbf{0.84} \\
 & CNN & 5 & 20 & 84.8 & 83.0 & 83.9 & 0.678 \\
 & CNN & 10 & 20 & 88.0 & 88.2 & 88.1 & 0.761 \\
 & CNN & 20 & 20 & 81.7 & 91.4 & 91.0 & 0.817 \\
 \hline
 \end{tabular}
 }
 \end{table}
From all experiments about Taiwan 50, we conclude a summary result with and without volume indicator for different trading days’ period and image dimension result. Table \ref{Tab:summarytw50} shows that CNN in 20 trading days’ period with 50-dimension image and volume indicator is better than the others with 91.5\% accuracy. In addition, without volume indicator for Taiwan 50, CNN in 20 trading days’ period with 50 dimension performs better than the others with 92.2\% accuracy. From the result of both of those experiments, it indicates that the method using CNN model with longer trading day’s period without volume indicator can achieve the best result for Taiwan 50 dataset.
\subsection{Classification for Indonesia 10 Dataset}
\begin{table}[H]
 \centering
 \caption{Summary result of Indonesia 10 with their best classifier for each trading days and image dimension.}
 \label{Tab:summaryid10}
\scalebox{0.75}{%
 \begin{tabular}{|c|l|c|c|c|c|c|c|c|}
 \hline
 & \multicolumn{1}{c|}{Classifier} & \multicolumn{1}{c|}{Period} & \multicolumn{1}{c|}{Dimension} & \multicolumn{1}{c|}{Sensitivity} & \multicolumn{1}{c|}{Specitivity} & \multicolumn{1}{c|}{Accuracy} & \multicolumn{1}{c|}{MCC}\\
 \hline
 \parbox[t]{2mm}{\multirow{5}{*}{\rotatebox[origin=c]{90}{with volume}}} 
 & ResNet50 & 5 & 50 & 80.7 & 85.4 & 83.1 & 0.661 \\
 & ResNet50 & 10 & 50 & 88.6 & 88.4 & 88.5 & 0.77 \\
 & CNN & 20 & 50 & \textbf{90.0} & \textbf{90.1} & \textbf{90.0} & \textbf{0.798} \\
 & ResNet50 & 5 & 20 & 78.8 & 82.3 & 80.6 & 0.612 \\
 & CNN & 10 & 20 & 83.3 & 85.4 & 84.3 & 0.686 \\
 & CNN & 20 & 20 & 89.1 & 84.6 & 87.1 & 0.738 \\
 \hline
  \parbox[t]{2mm}{\multirow{5}{*}{\rotatebox[origin=c]{90}{without volume}}} 
 & ResNet50 & 5 & 50 & 79.1 & 87.9 & 83.3 & 0.671 \\
 & CNN & 10 & 50 & 87.5 & 86.6& 87.1 & 0.74 \\
 & CNN & 20 & 50 & \textbf{92.1} & 92.1 & \textbf{92.1} & \textbf{0.837} \\
 & CNN & 5 & 20 & 83.4 & 82.4 & 82.9 & 0.658 \\
 & CNN & 10 & 20 & 85.4 & 85.6 & 85.5 & 0.708 \\
 & VGG16 & 20 & 20 & 91.5 & 89.7 & 90.7 & 0.808 \\
 \hline
 \end{tabular}
 }
 \end{table}
From all experiment results with Indonesia 10 dataset, we conclude a summary result with and without volume indicator in Table \ref{Tab:summaryid10} respectively. It shows that the CNN method with 20 trading days’ period in 50 dimension using volume indicator show the best result with 90.0\% accuracy. While the CNN method in 20 trading days’ period with 20-dimension image without using the volume indicator performs better result with 92.1\% accuracy. It indicates that the method using CNN model with longer trading day’s period without volume indicator can achieve the best result for Indonesia 10 dataset.
\subsection{Independent Testing Result}
 \begin{table}[H]
 \centering
 \caption{Summary result of Taiwan 50 with their best classifier for each trading days and image dimension.}
 \label{Tab:indptw50}
\scalebox{0.75}{%
 \begin{tabular}{|c|l|c|c|c|c|c|c|c|}
 \hline
 & \multicolumn{1}{c|}{Classifier} & \multicolumn{1}{c|}{Period} & \multicolumn{1}{c|}{Dimension} & \multicolumn{1}{c|}{Sensitivity} & \multicolumn{1}{c|}{Specitivity} & \multicolumn{1}{c|}{Accuracy} & \multicolumn{1}{c|}{MCC}\\
 \hline
 \parbox[t]{2mm}{\multirow{5}{*}{\rotatebox[origin=c]{90}{with volume}}} 
 & CNN & 5 & 50 & 83.2 & 83.8 & 83.5 & 0.67 \\
 & CNN & 10 & 50 & 88.6 & 87.3 & 88.0 & 0.758 \\
 & CNN & 20 & 50 & \textbf{91.6} & \textbf{91.3} & \textbf{91.5} & \textbf{0.827} \\
 & CNN & 5 & 20 & 83.9 & 82.7 & 83.3 & 0.666 \\
 & Random Forest & 10 & 20 & 87.0 & 88.3 & 87.6 & 0.751 \\
 & CNN & 20 & 20 & 90.8 & 90.2 & 90.6 & 0.808 \\
 \hline
  \parbox[t]{2mm}{\multirow{5}{*}{\rotatebox[origin=c]{90}{without volume}}} 
 & CNN & 5 & 50 & 83.6 & 85.1 & 84.4 & 0.687 \\
 & CNN & 10 & 50 & 89.2 & 88.1 & 88.7 & 0.773 \\
 & CNN & 20 & 50 & \textbf{93.3} & 90.7 & \textbf{92.2} & \textbf{0.84} \\
 & CNN & 5 & 20 & 84.8 & 83.0 & 83.9 & 0.678 \\
 & CNN & 10 & 20 & 88.0 & 88.2 & 88.1 & 0.761 \\
 & CNN & 20 & 20 & 81.7 & 91.4 & 91.0 & 0.817 \\
 \hline
 \end{tabular}
 }
 \end{table}
  \begin{table}[H]
 \centering
 \caption{Summary result of Taiwan 50 with their best classifier for each trading days and image dimension.}
 \label{Tab:indpid10}
\scalebox{0.75}{%
 \begin{tabular}{|c|l|c|c|c|c|c|c|c|}
 \hline
 & \multicolumn{1}{c|}{Classifier} & \multicolumn{1}{c|}{Period} & \multicolumn{1}{c|}{Dimension} & \multicolumn{1}{c|}{Sensitivity} & \multicolumn{1}{c|}{Specitivity} & \multicolumn{1}{c|}{Accuracy} & \multicolumn{1}{c|}{MCC}\\
 \hline
 \parbox[t]{2mm}{\multirow{5}{*}{\rotatebox[origin=c]{90}{with volume}}} 
 & CNN & 5 & 50 & 83.2 & 83.8 & 83.5 & 0.67 \\
 & CNN & 10 & 50 & 88.6 & 87.3 & 88.0 & 0.758 \\
 & CNN & 20 & 50 & \textbf{91.6} & \textbf{91.3} & \textbf{91.5} & \textbf{0.827} \\
 & CNN & 5 & 20 & 83.9 & 82.7 & 83.3 & 0.666 \\
 & Random Forest & 10 & 20 & 87.0 & 88.3 & 87.6 & 0.751 \\
 & CNN & 20 & 20 & 90.8 & 90.2 & 90.6 & 0.808 \\
 \hline
  \parbox[t]{2mm}{\multirow{5}{*}{\rotatebox[origin=c]{90}{without volume}}} 
 & CNN & 5 & 50 & 83.6 & 85.1 & 84.4 & 0.687 \\
 & CNN & 10 & 50 & 89.2 & 88.1 & 88.7 & 0.773 \\
 & CNN & 20 & 50 & \textbf{93.3} & 90.7 & \textbf{92.2} & \textbf{0.84} \\
 & CNN & 5 & 20 & 84.8 & 83.0 & 83.9 & 0.678 \\
 & CNN & 10 & 20 & 88.0 & 88.2 & 88.1 & 0.761 \\
 & CNN & 20 & 20 & 81.7 & 91.4 & 91.0 & 0.817 \\
 \hline
 \end{tabular}
 }
 \end{table}
Measuring our model result not only used performance evaluation. We also performed an independent test to see that our proposed method is reasonable. During this independent test, we used two index stock exchange data from each country. Yuanta/P-shares Taiwan Top 50 ETF represented independent data test for our Taiwan50, whereas Jakarta Composite Index is our independent data set test for Indonesia10. Both of the stock exchange data are taken from 1st January, 2017 until 14th June 2018. 
Table \ref{Tab:indptw50} shows our independent test result for Taiwan50 using volume indicator and without using volume indicator respectively. The independent test result for Indonesia10 using and without using volume indicator are shown in Table \ref{Tab:indpid10} respectively. As shown in Tables \ref{Tab:indptw50} and \ref{Tab:indpid10}, our CNN with 20 trading days’ period and 50-dimension image get best result for both independent test.
\subsection{Comparison}
  
To further evaluate the effectiveness of our predictive model, we also compare our result with the other related works. The first comparison is between our proposed method with Khaidem's work\cite{khaidem2016predicting}, they used three different stock market datasets with different trading period time. Samsung, General Electric and Apple are their stock market data with one, two and three months of trading period respectively. We applied our proposed model in their datasets to compare our prediction performance with their result. The comparison result for Samsung, Apple, and GE stock market shown in Table \ref{Tab:khaidemtab} respectively. Based on these comparison results, it revealed that our performance results outperformed the prediction results from Khaidem's work\cite{khaidem2016predicting}.
\begin{table}[H]
 \centering
 \caption{Comparison result with Khaidem.}
 \label{Tab:khaidemtab}
\scalebox{0.8}{%
\begin{tabular}{|l|c|c|c|c|c|}
\hline
\multicolumn{6}{|c|}{Khaidem, Saha et al. – Samsung}                                                         \\ \hline
\multicolumn{1}{|c|}{Name} & Trading Period & ACC           & Precision     & Recall        & Specificity    \\ \hline
Khaidem                    & 1 Month        & 86.8          & 88.1          & 87.0          & 0.865          \\ \hline
Our                        & 1 Month        & \textbf{87.5} & 88.0          & \textbf{87.0} & \textbf{0.891} \\ \hline
Khaidem                    & 2 Month        & 90.6          & 91.0          & 92.5          & 0.88           \\ \hline
Our                        & 2 Month        & \textbf{94.2} & \textbf{94.0} & \textbf{94.0} & 0.862          \\ \hline
Khaidem                    & 3 Month        & 93.9          & 92.4          & 95.0          & 0.926          \\ \hline
Our                        & 3 Month        & \textbf{94.5} & \textbf{94.0} & \textbf{95.0} & 0.882          \\ \hline
\multicolumn{6}{|c|}{Khaidem, Saha et al. – Apple}                                                           \\ \hline
Khaidem                    & 1 Month        & 88.2          & 89.2          & 90.7          & 0.848          \\ \hline
Our                        & 1 Month        & \textbf{89.6} & \textbf{90.0} & 90.0          & \textbf{0.863} \\ \hline
Khaidem                    & 2 Month        & 93.0          & 94.1          & 93.8          & 0.919          \\ \hline
Our                        & 2 Month        & \textbf{93.6} & 94.0          & \textbf{94.0} & 0.877          \\ \hline
Khaidem                    & 3 Month        & 94.5          & 94.5          & 96.1          & 0.923          \\ \hline
Our                        & 3 Month        & \textbf{95.6} & \textbf{96.0} & 96.1          & 0.885          \\ \hline
\multicolumn{6}{|c|}{Khaidem, Saha et al. – GE}                                                              \\ \hline
Khaidem                    & 1 Month        & 84.7          & 85.5          & 87.6          & 0.809          \\ \hline
Our                        & 1 Month        & \textbf{90.2} & \textbf{90.0} & 90.0          & 0.86           \\ \hline
Khaidem                    & 2 Month        & 90.8          & 91.3          & 93.0          & 0.876          \\ \hline
Our                        & 2 Month        & \textbf{97.8} & \textbf{98.0} & \textbf{98.0} & \textbf{0.993} \\ \hline
Khaidem                    & 3 Month        & 92.5          & 93.1          & 94.5          & 0.895          \\ \hline
Our                        & 3 Month        & \textbf{97.4} & \textbf{98.0} & \textbf{98.0} & \textbf{0.983} \\  \hline
 \end{tabular}
 }
 \end{table}
  \begin{table}[H]
 \centering
 \caption{Comparison result with Patel.}
 \label{Tab:pateltab}
\scalebox{0.8}{%
\begin{tabular}{|c|c|c|c|c|}
\hline
\multicolumn{3}{|c|}{S\&P BSE SENSEX}   & \multicolumn{2}{c|}{NIFTY 50} \\ \hline
       & ACC            & F-Measure     & ACC           & F-Measure     \\ \hline
Patel  & 89.84          & 0.9026        & 89.52         & 0.8935        \\ \hline
Our    & \textbf{97.2}  & \textbf{0.97} & \textbf{93.4} & \textbf{0.93} \\ \hline
\multicolumn{3}{|c|}{Reliance Industry} & \multicolumn{2}{c|}{Infosys}  \\ \hline
Patel  & 92.22          & 0.9234        & 90.01         & 0.9017        \\ \hline
Our    & \textbf{93.9}  & \textbf{0.94} & \textbf{93.9} & \textbf{0.94} \\ \hline
 \end{tabular}
 }
 \end{table}
Second comparison is between our proposed method with J. Patel's work\cite{patel2015predicting}. They utilized four different stock market datasets from India stock exchange. In this comparison, we followed their dataset using Nifty50, S7P BSE Sensex, Reliance Industry and Infosys stock market datasets. Accuracy and F-measure were used for their performance evaluation. As comparison results shown in Table \ref{Tab:pateltab}, Our proposed model yielded 97.2 \%, 93.9 \%, 93.4 \% and 93.9 \% for accuracy with S7P BSE Sensex, Reliance Industry, Nifty50 and Infosys stock market datasets respectively. It indicated that our proposed method is superior to Patel work \cite{patel2015predicting}.
  \begin{table}[H]
 \centering
 \caption{Comparison result with Zhang.}
 \label{Tab:zhangtab}
\scalebox{0.8}{%
\begin{tabular}{|l|c|c|}
\hline
\multicolumn{3}{|c|}{Hong Kong - Zhang} \\ \hline
\multicolumn{1}{|c|}{} & Accuracy & MCC \\ \hline
Zhang & 61.7 & 0.331 \\ \hline
Our & \textbf{92.6} & \textbf{0.846} \\ \hline
 \end{tabular}
 }
 \end{table}
Last comparison is between our proposed method with Zhang's method\cite{zhang2018improving}. Their dataset composition is similar with us. They are using thirteen Hong Kong stock market, whereas we used fifty Taiwan stock market datasets and ten Indonesia stock market datasets. Their methodology is combine sentiment analysis on social media and finance news. As shown in Table \ref{Tab:zhangtab}, Our proposed method achieved 92 \% significantly outperforms Zhang method \cite{zhang2018improving}.
\section{Conclusions and Future Works}

In this study, we present a new method for stock market prediction using 2 stock market datasets including 50 company stock markets for Taiwan50 datasets and 10 company stock market for Indonesian datasets. The first, we employ the sliding window technique to generate the period data. To find out correlation between enrich candlestick chart information and stock market prediction performance, we utilized the computer graphic technique to generate the candlestick chart images for stock market data. Finally, an CNN learning algorithm is employed to build our prediction for stock market.
\par
We found that the model using long-term trading days’ period with CNN learning algorithm achieves the highest performance of sensitivity, specificity, accuracy, and MCC. It is proved that Convolutional neural network can find the hidden pattern inside the candlestick chart images to forecast the movement of specific stock market in the future. Adding the indicator such as volume in candlestick chart not really help the algorithms increase finding the hidden pattern.
\par
The comparison experiments indicated that our proposed method provide highly accurate forecast for other datasets compare to the other existing methods. Patel used trading data from Reliance Industries, Infosys Ltd., CNX Nifty and S \& P Bombay Stock Exchange BSE Sensex during 10 years with accuracy in the range of 89 \% - 92 \% while we achieved accuracy in the range of 93 \% - 97 \%. Khaidem method achieved the accuracy in the range of 86 \% - 94 \% using three trading data from Samsung, GE and Apple while we achieved in the range of 87 \% - 97 \%. Zhang utilized 13 different companies in Hong Kong stock exchange with accuracy 61 \%. Meanwhile, our method achieved 92 \% for accuracy. 
\par
For the future works we want to extend our work being able to predict the percentage change on the price movements. For the convenience of experimental scientists, we developed a user-friendly webserver for predicting stock market using our final model. Available at \url{http://140.138.155.216/deepcandle/}, DeepCandle is a system through which users can easily predicting stock market in the near future. Users only need to input the target date, and our models will process them and return the prediction result of stock market movement on that target date. The provided web interface is constructed such that users can easily access its functions and comfortably use it without a deep understanding of computing.

\bibliographystyle{abbrv}
\bibliography{refs}
\end{document}